\documentstyle{pasj00}
\begin{document}

\SetRunningHead{A.Imada \& B.Monard}{Dwarf Nova ASAS 160048-4846.2}

\title{Discovery of a Promissing Candidate of WZ Sge-Type Dwarf Novae,
ASAS 160048-4846.2: Evidence for Double-Peaked Humps}

\author{Akira \textsc{Imada}$^1$ and L.A.G. Berto \textsc{Monard}$^{2}$}

\affil{$^1$Department of Astronomy,Faculty of Science, Kyoto University,
       Sakyo-ku, Kyoto 606-8502}
\affil{$^2$Bronberg Observatory, PO Box 11426, Tiegerpoort 0056, South
       Africa} 
\email{a\_imada@kusastro.kyoto-u.ac.jp}

\KeyWords{
          accretion, accretion disks
          --- stars: dwarf novae
          --- stars: individual (ASAS 160048-4846.2)
          --- stars: novae, cataclysmic variables
          --- stars: oscillations
}

\maketitle

\begin{abstract}
We report on time-resolved CCD photometry during the 2005 June outburst
 of a dwarf nova, ASAS160048-4846.2. The observed light curves
 unambiguously showed embryonic humps with a period of 0.063381(41)
 days, after which genuine superhumps emerged with a period of
 0.064927(3) days. Based on evidence for double-peaked humps in the
 earlier stage of the outburst, this object might be qualified as
 the seventh member of WZ Sge-type dwarf novae after Var Her 04. If the
 former period is the same as, or very close to the orbital period of
 the system, as in other WZ Sge systems, the fractional
 superhump excess is about 2.4${\%}$. This value is unexpectedly larger
 than that of other WZ Sge-type dwarf novae. The early phase of our
 observing run provided evidence for the transition from chaotic
 humps to genuine superhumps, together with increasing the amplitude. 
\end{abstract}

\section{introduction}

Recently, extensive monitoring of dwarf novae has revealed that SU
UMa stars have diversity of the behavior. WZ Sge stars, an extreme
subclass of SU UMa stars, exhibit peculiarities compared
to well-observed SU UMa stars \citep{kat01hvvir}. Main properties
of WZ Sge stars are that (1) they have very long recurrence times,
sometimes in excess of decades, (2) the amplitude of these systems
exceeds 6 mag instead about 4${\sim}$5 mag for many
SU UMa stars, (3) one outburst or repetitive ones take place after the
termination of the main superoutburst, and (4) double-peaked humps, with a
periodicity of almost the same as the orbital period of the system
(\cite{osa02wzsgehump}; \cite{pat02wzsge}; \cite{kat02wzsgeESH}),
are observed on the early phase of superoutbursts. Especially, the
last one is of particular interest because there are only 6 systems that
showed double-peaked humps: AL Com (\cite{kat96alcom},
\cite{pat96alcom}, EG Cnc \citep{pat98egcnc}, RZ Leo \citep{ish01rzleo},
HV Vir \citep{ish03hvvir}, Var Her 04 \citep{pri04varher04}, and WZ Sge
itself \citep{kat04vsnet}. 

There are three competing models concerning the double-peaked humps
at the early phase of the superoutburst for WZ Sge
stars since the 2001 superoutburst of WZ Sge
itself. \citet{osa02wzsgehump} proposed that the double-peaked humps are
most likely a manifestation of the 2:1 resonance radius in the accretion
disk in these systems with extremely low mass
ratios. \citet{kat02wzsgeESH} suggested that irradiation of the elevated
surface of the accretion disk caused by vertical tidal deformation well
represents the observations. On the other hand, \citet{pat02wzsge}
proposed that enhanced-mass transfer from the secondary causes more
luminous hot-spot on the accretion disk, resulting in double-peaked
humps The presence of the double-peaked humps is now regarded as a
criterion for WZ Sge stars \citep{ima05gocom}. Many works have been
performed both from theoretical and observational sides in order to
decipher the nature of double-peaked humps, as well as other properties
(\cite{las95wzsge}; \cite{osa03DNoutburst}. However, there are many
problems left behind our understanding.

ASAS 160048-4846.2 (hereafter ASAS 1600) was discovered as an eruptive
object by All Sky Automated Survey \citep{poj02asas3} on June 9
2005. This is the first recorded superoutburst in the survey. Astrometry
on the UCAC2 frames yields the coordinate of the variable with
$16^{\rm h} 00^{\rm m} 47^{\rm s}.43$, $-48^{\circ} 46' 07''.6$.

In this letter, we mainly focus on the variation of hump 
profiles. Detailed analysis will be published in the forthcoming paper.

\section{observation}

Time-resolved CCD photometries were performed with a 32 cm telescope at
 Tiegerpoort (South Africa) for 12 consecutive nights between 2005 June
 9 and June 20. The exposure time during the run was 30 seconds. The
 total data points amounted to 10511. After dark subtraction and flat
 fielding on the acquisitions, we analyzed the target, a comparison
 star, and a check star by aperture photometry. For a comparison star
 and check star, we chose UCAC2 160053.1-484433 ($R$ = 11.9) and UCAC2
 160043.6-484628 ($R$ = 12.8), respectively. Heliocentric corrections to
 the observed times were applied before the following analysis.

\section{results}

\subsection{light curve}

Figure 2 shows the whole light curves obtained during our run. The plateau
stage lasted more than 12 days with a declining rate of 0.12 mag
d$^{-1}$. During a plateau stage, some WZ Sge stars show steep decline
just after the maximum at a rate of 0.5 mag d$^{-1}$ (e.g., WZ Sge), or
a brightening up to 0.2 mag at the middle of the plateau
phase (e.g., EG Cnc, \citep{mat98egcnc}). As for ASAS 1600, such
features could not be detected during our run.

After the termination of the plateau phase, ASAS 1600 underwent a
rebrightening outburst on HJD 2453547 detected by R. Stubbings
(vsnet-outburst 6503) with $V$ ${\sim}$ 14.5. We have no other
information about the rebrightening.

\begin{figure}
\begin{center}
\resizebox{80mm}{!}{\includegraphics{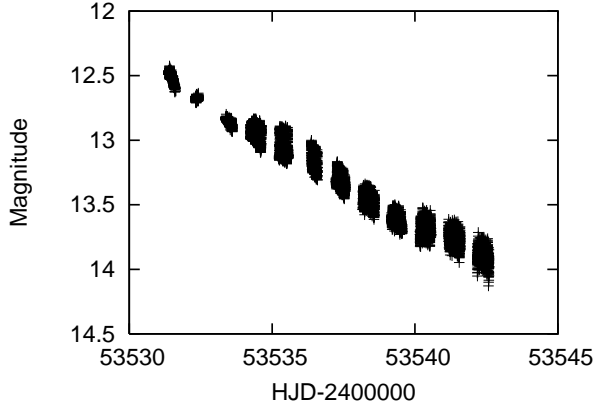}}
\end{center}
\caption{Light curves during the plateau stage. The vertical and
 horizontal axes denote the $R$ band magnitude and HJD,
 respectively. The light curves also showed a linear decline trend at a
 rate of 0.12 mag d$^{-1}$, on which modulations with an amplitude of
 about 0.2 mag were superimposed.}
\end{figure}

\subsection{superhumps}

Superhump features were prominent from HJD 2453534, the fourth day of
our run. After subtracting a declining trend of the plateau stage, we
carried out the PDM method \citep{ste78pdm} from HJD 2453534 to HJD
2453542. Figure 2 represents the theta diagram of this stage, from which
we can conclude that the mean superhump period is 0.064927(3) days with
1-${\sigma}$ error. The error was calculated by the Lafler-Kinman
methods \citep{fer89error}. Notice that this period is much longer than
those of known WZ Sge stars, e.g., 0.05721 days for AL Com
\citep{nog97alcom}, 0.05820 days for HV Vir \citep{ish03hvvir}. This
value is quite similar to that of a promising candidate of WZ Sge star,
ASAS 153616-0837.1 \citep{pat05re1255}. No other periodicity was found
during the span.

Phase-averaged light curves folded by 0.064927 days are depicted in
figure 3, in which one can see the ``textbook'' feature of
superhumps with an amplitude of about 0.2 mag. There is no sign of an
eclipse, indicating a low inclination of the system. 

\begin{figure}
\begin{center}
\resizebox{80mm}{!}{\includegraphics{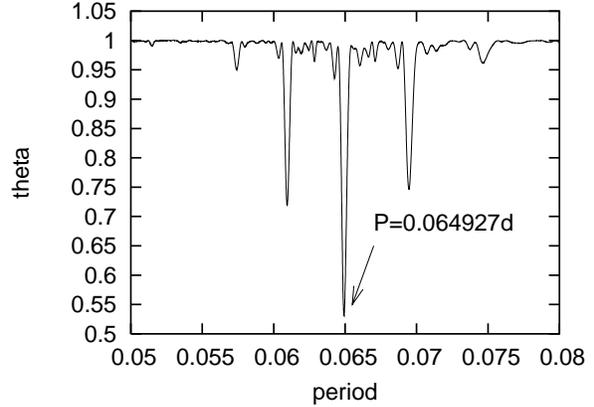}}
\end{center}
\caption{Period analysis of genuine superhumps in ASAS 1600-48. Data of
 the first three days (from HJD 2453531 to HJD 2453533) were precluded
 during the analysis because the profiles during the stages were
 ambiguous. We determined 0.064927(3) days as the best superhump
 period.}
\end{figure}

\begin{figure}
\begin{center}
\resizebox{78mm}{!}{\includegraphics{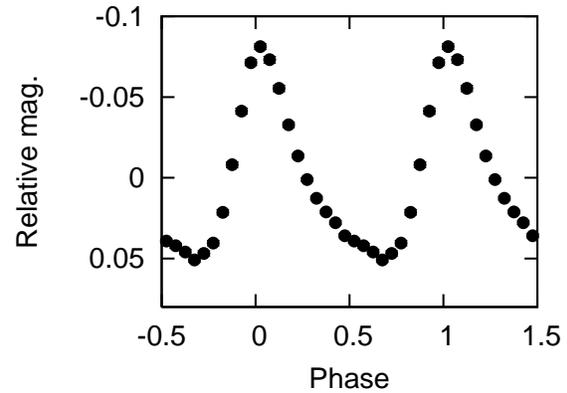}}
\end{center}
\caption{Mean superhump profile between HJD 2453534 and HJD 2453542,
 folded by 0.064927 days, when conspicuous modulations were shown. The
 vertical and horizontal axes show the relative magnitude and phase,
 respectively. The epoch of the phase was set on the time of the first
 superhump maximum in our observations. There provided no evidence for
 an eclipse, suggesting a low - medium inclination of the system.}
\end{figure}

\subsection{early stage of the outburst}

\begin{figure*}
\begin{center}
\resizebox{53mm}{!}{\includegraphics{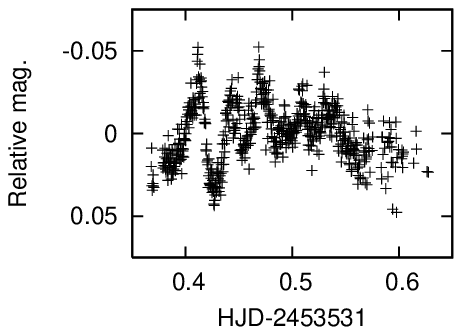}}
\resizebox{53mm}{!}{\includegraphics{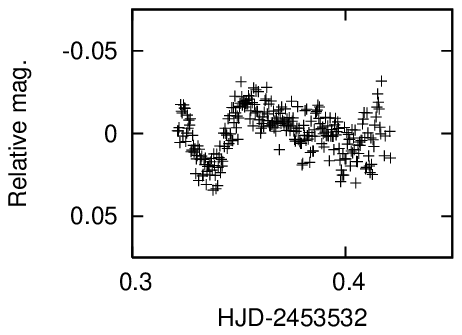}}
\resizebox{53mm}{!}{\includegraphics{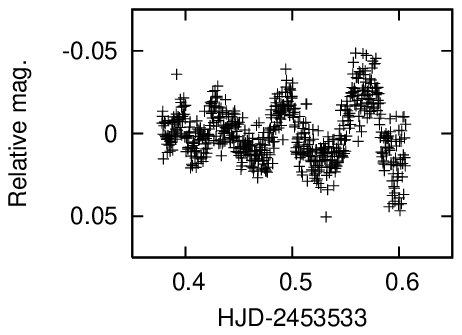}}
\end{center}
\caption{Enlarged light curves on the first 3 days of our
 run. On HJD 2453533, one can see four humps in the figure. Note, that
 the amplitude of the hump increased with time, but the less than 0.1
 mag, suggesting we are the witness of the growth of superhumps. }
\end{figure*}

At the earlier stage of the superoutburst, the light curve showed
ambiguous humps of variable shapes. Judging from our run, such
modulations lasted 3 days. Figure 4 represents enlarged light curves of
this phase. In order to search for a periodicity, we performed the PDM
method \citep{ste78pdm} over the first two days after subtracting the
declining trend. Figure 5 demonstrates the theta diagram for the search,
for which we found a periodicity of 0.063381(41) days. This period is
certainly not an alias, but the period of this stage. As is often
observed in other WZ Sge stars, this period might be the same as, or very
close to the orbital period of the system. However, spectroscopic
observations for the object should be performed to confirm this.

We also made phase-averaged light curves folded by the period (figure
6). Note that the profile fairly represents a double-peaked
feature as is observed in other WZ Sge stars. This results could allow us
to suggest that ASAS 1600 may be a promissing candidate for WZ Sge-type
dwarf novae.

The right panel of figure 4 shows the light curves on HJD 2453533 in
which we detected the transition from ambiguous humps to genuine
superhumps. Although the amplitude of the humps is as less as 0.1 mag,
it can be obviously seen the growing humps.

\begin{figure}
\begin{center}
\resizebox{80mm}{!}{\includegraphics{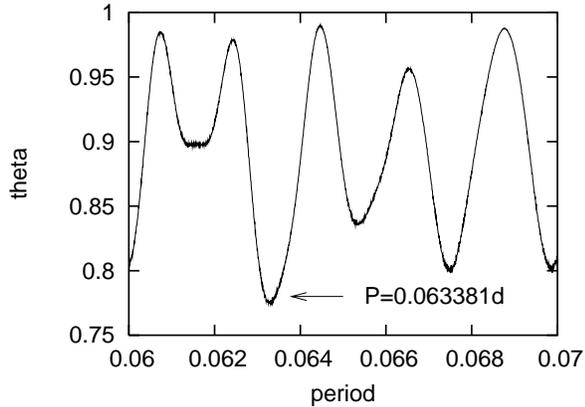}}
\end{center}
\caption{Period search for the first 2 days of the observations. The
 theta diagram show the periodicity of 0.063381(41) days, which we
 interpret as the period of early superhumps seen in WZ Sge-type dwarf
 novae.}
\end{figure}

\begin{figure}
\begin{center}
\resizebox{80mm}{!}{\includegraphics{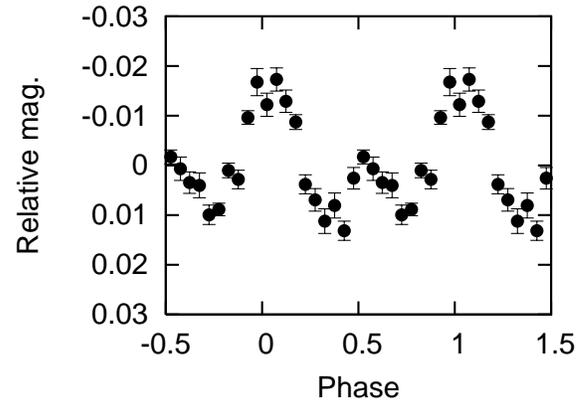}}
\end{center}
\caption{Mean humps on the first 2 days of our run, folded by 0.063381
 days. The abscissa and the ordinate denote the phase and relative
 magnitude, respectively. The epoch was arbitrary set. Note, that the
 profiles are not single-peaked, but double-peaked. Such features are
 characteristic of WZ Sge stars on the earlier stage of superoutburst.}
\end{figure}

\section{discussion}

As mentioned above, WZ Sge stars rarely exhibit a superoutburst, which has
hampered us from qualification of a new member of the systems, although a
possible candidate for WZ Sge stars amounts to a few tens
(\cite{kat01hvvir}; \cite{szk03CVSDSS}). Our observations proved that
ASAS 160048-4846.2 is the seventh member of WZ Sge stars after
detection of double-peaked humps and a rebrightening. The outburst
properties were similar to a long-period WZ Sge star, RZ Leo, in terms
of profiles of humps features. 

As for double-peaked humps at the early stage of the superoutburst, we
detected a periodicity of 0.063381(41) days. This period is slightly
shorter than the superhump period of 0.064927(3) days. $If$ we could
interpret the former period as the orbital period of the system, the
fractional superhump excess is 0.024(1), and we can roughly estimate the
mass ratio of the system using an empirical relation as follows
\citep{pat98evolution},

\begin{equation}
\epsilon = \frac{0.23q}{1+0.27q},
\end{equation}

where ${\epsilon}$ and $q$ are the fractional superhump excess and mass
ratio, respectively. With a little algebra, we can determine the value
of $q$ with 0.109(4). Compared to other WZ Sge stars, this value is
slightly large. In conjunction with the obtained values, it is
likely that the secondary of ASAS 1600 is not a degenerated brown
dwarf, but a normal M-type star. \citep{pat05re1255} recently suggests
that a possible dwarf nova, RE 1255+266, may be a promising candidate
for ``period bouncer'', whose secondary star has a very low mass and is
degenerated. \citet{pat05re1255} further derived that the mass ratio of
RE 1255+266 to be less than 0.06. If the obtained period of
double-peaked humps does indeed correspond to the orbital period of the
system, we might exclude the possibility that ASAS 1600 is ``period
bouncer''. This also implies that WZ Sge stars are not restricted to
systems having a low mass secondary star, which is strongly supported by
the observations of RZ Leo, the longest period of WZ Sge stars ever
known. The mass ratio of RZ Leo is estimated to be 0.14 using the above
equation \citep{ish01rzleo}, indicating a normal secondary of the
object. This is consistent with a spectroscopic observation by
\citet{men99rzleo}.

Spectroscopic observations should be performed in order to determine the
precise orbital period of the system, which will yield the accurate
value of superhump excess.

We also detected the transition from double-peaked humps to genuine
superhumps on HJD 2453533. This phenomena are difficult to observe
because this may rapidly occur. During this, the magnitude of systems
increases by ${\sim}$0.2 mag, which is observed in EG Cnc
\citep{mat98egcnc}, AL Com (\cite{pat96alcom}; \cite{how96alcom}) Var
Her 04 \citep{pri04varher04}. On the other hand, ASAS 1600 showed no
resumption of the magnitude \citep{pat02wzsge}. So far, we cannot
specify the reason why EG Cnc and Var Her 04 took place and ASAS 1600
did not. Further observations during the transition will improve our
understanding of this phenomenon.

\vskip 3mm

We express our gratitude to Daisaku Nogami for his careful reading and
constructive suggestions on the first draft.

\end{document}